# Classical isotropic two body potentials generating martensitic transformations

M F Laguna and E A Jagla

Centro Atómico Bariloche and CONICET, Bustillo 9500 (8400) Bariloche, Río Negro, Argentina

E-mail: lagunaf@cab.cnea.gov.ar and jagla@cab.cnea.gov.ar

**Abstract.** An isotropic interaction potential for classical particles is devised in such a way that the crystalline ground state of the system changes discontinuously when some parameter of the potential is varied. Using this potential we model martensitic transformations, and are able to study in detail the processes that are usually associated with it: shape memory effect, superelasticity, as well as many details concerning the dynamics of the transformation, particularly the characteristics of the martensitic texture obtained as a function of parameters affecting the transformation rate. Here we introduce the interaction potentials and present some basic results about the transformation it describes, for the particular case of two dimensional triangular-rombohedral and triangular-square transformation.

## 1. Introduction

A martensitic transformation is a solid state non-diffusive first order transition between a parent (usually higher temperature) crystalline phase called the austenite, and a lower symmetry (lower temperature) phase called the martensite. In this transformation the displacement of individual atoms with respect to its neighbors is small, comparable to (and typically much lower than) the interatomic distance itself. When a univocous correspondence exists between the atomic positions of the martensite and those of the austenite, the material displays the notorious phenomenon of shape memory that has many practical applications as well as very interesting theoretical aspects [1, 2].

Central to the shape memory effect and to the phenomenology of martensites in general is the fact that from a single crystal austenite, the martensite phase can be obtained with different orientations. They are termed the martensitic variants and can be thought as originated in distortions of the austenite structure along different (but crystallografically equivalent) directions when the martensite is formed. In two dimensions the austenite phase is typically a hexagonal (triangular) structure and the martensitic variants will be associated with three equivalent deformations along the three sides of the triangle. In figure 1(a) we schematically show this situation, where the deformation consists on an elongation of one of the sides and the contraction of the other two in such a way that the volume remains almost constant. Another important concept in martensitic transformations is that of a habit or invariant plane, or in two dimensions a habit line. This is a line along which martensite and austenite can match together, without generating long range elastic distortions in the system. In three dimensions, it turns out that a habit plane between austenite and a single variant of martensite does not exist, in general [2]. A mixture of two martensite variants is needed to obtain a habit plane. In the simpler two dimensional case, habit lines exist between austenite and single variant martensite. An example is shown in figure 1(b).

Experimental systems displaying martensitic transformations are usually alloys of two or more metallic elements, although there are also ceramic materials having this property [1]. In these systems, the transformation usually occurs due to entropic effects as the temperature is changed. This is at first sight striking, since thermal diffusion is negligible during martensitic transformations. One can recognize however, that the role of temperature can be represented (having in mind a Ginsburg-Landau description [3]) at two different levels. On one hand we may have an effective dependence of the parameters of the system on temperature, and in addition to that, we can have stochastic effects of temperature, producing in particular atom diffusion. It is the latter effect that is negligible in martensitic transformations, but the first does not need to be so.

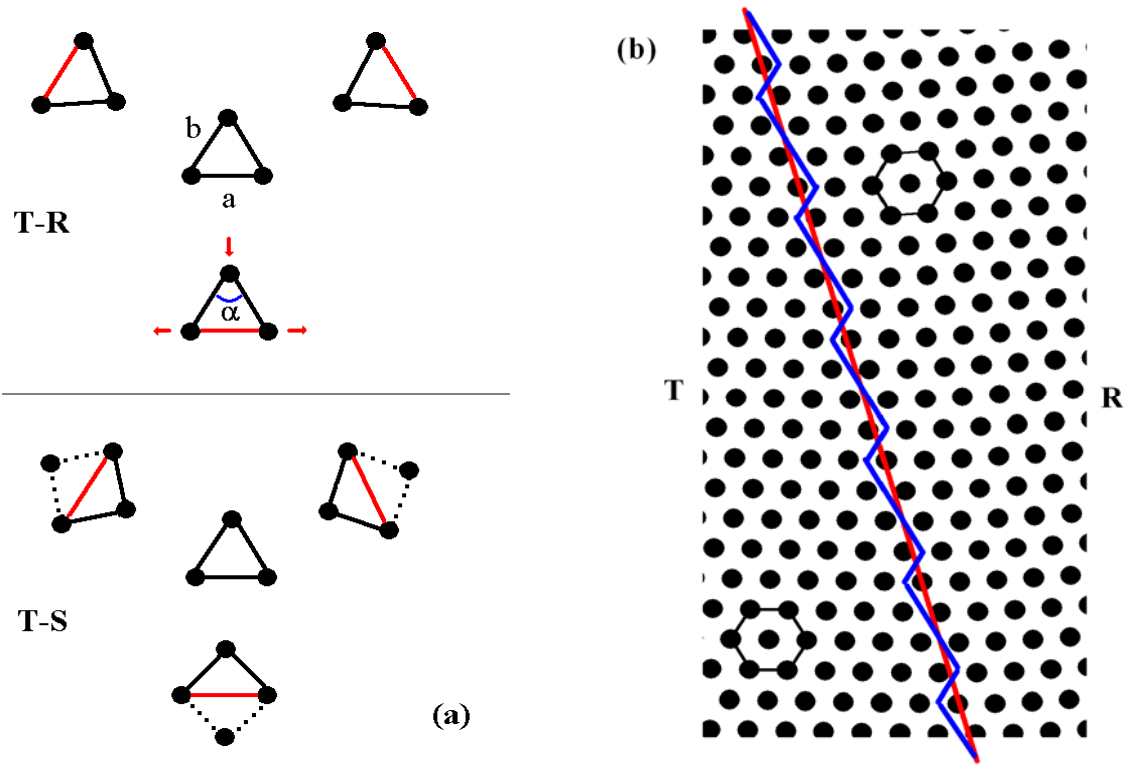

Figure 1: **(a)** Martensitic variants in two dimensions for the triangular-rombohedral (T-R) transition and triangular-square (T-S) transition. Red segment indicates the most elongated side of the triangle. The angle α is less than π/2 in the first case, and equal to π/2 in the second. **(b)** Habit line (in red) separating the triangular (T) region from the rombohedral (R) one. The zigzag line separates individual particles corresponding to both phases.

In fact, to obtain a very simplified description of the martensitic transformation, we can imagine a system in its ground state in which we change the interparticle potential in a prescribed form, in such a way that a stability change between two competing structures occurs.

The possibility of a system to exhibit a martensitic transformation is usually associated with the existence of different kind of particles interacting through rather complex (for instance, angle dependent) potentials. In a recent attempt to get a simpler realization, Rao *et. al* have used an effective potential constructed as a sum of a repulsive anisotropic two body potential and a short range three body potential to study solid state transformations of the martensitic type [4,5].

However, it is becoming widely recognized that is feasible to have different crystalline structures for identical particles interacting with specially devised two-body isotropic potentials [6,7]. Now it is clear that isotropic potentials can be devised to obtain a variety of structures. To our knowledge the techniques of designing interaction potentials have not been used before to obtain spherically symmetric, two body classical potentials displaying martensitic transformations, either in two or three dimensions.

We demonstrate here that simple interparticle potentials in single-species systems are sufficient to realistically study the phenomena associated with martensitic transformations.

In the present paper we introduce the kind of potentials giving rise to martensitic transformations, and discuss some associated phenomenology that is reproduced with our model (see also Ref. [8]).

The particular transformations we have studied are the 2D triangular-rombohedral, rombohedral-square, and triangular-square transformations. In 3D we have potentials describing fcc-bct, and fcc-bcc transformations. Other possibilities appear but we have not studied them in detail yet.

For clarity of presentation we restrict in the present paper to the 2D transformations, studying first the triangular-rombohedral (T-R) transformation, where a shape memory effect is expected, and compare with the case of the triangular-square (T-S) transformation, where shape memory effect is not expected

[9]. We leave for a forthcoming publication the study of the richer (but also more involved) case of full three dimensional transformations.

The paper is organized as follows: in the next section we give a description of the potentials that we have constructed, give some detail on the simulation method, and show the expected ground state structures. In Section 3 we show the results of simulations on a single crystal austenite phase, focusing on the effects that transformation rate and quench depth have on the final microstructure of the martensite. In Section 4 we apply an external stretching on the martensitic sample in order to study the shape memory effect. In Section 5 we discuss how the same protocol of Section 4, when applied to the triangular square transformation, fails to produce shape memory. Finally, in Section 6 we present our conclusions.

## 2. Details on the model

Realistic interatomic potentials are usually found using ab initio calculations, and are available for most of the elements. Usually, these kind of potentials are rather complex, involving in particular non trivial angular dependences of the interactions. The complexity of such interactions makes appealing the idea of devising simple potentials that can give a reasonable description of the problem of interest, being at the same time sufficiently simple to allow numerical or analytical calculations. Along this line, the use of spherically symmetric potential has a long history, being the 6-12 Lennard Jones potential the most widely known case. This potential was originally introduced to study the interactions between atoms of noble gases, and in particular, it always gives rise to compact ground state structures (fcc or bcc), and of course, to a triangular ground state structure in two dimensions.

It has not been recognized until recently, that isotropic potentials are able to generate also a large variety of ground state configurations other than compact ones, both in two and three dimensions [6,7,10]. In many cases this behavior can be obtained by slight variations of the potentials with respect to the Lennard Jones prototype (note that we restrict here to the case in which all pair of particles share the same interaction potentials; different structures can be more or less trivially obtained in the case in which particles have two or more different sizes).

We exploit this fact to construct potentials that produce ground state configurations other than the trivial ones. Moreover, by varying some potential parameters we can produce an interchange of stability between two of these phases and then produce a martensitic transformation.

Our work in finding such potential was based in the previous experience with potentials that display pressure induced amorphisation [11,12]. In those studies we had already indicated that some of the transitions we had found were very much like martensitic transformations. But beyond this motivation, our work has had a great deal of trial an error.

Our final potentials look qualitatively like the standard Lennard Jones one, with a hard core at short distances, and an attractive tail at large distances. However slight differences exist that modify the contributions to the energy of second and further neighbors, and this generates the possibility of other structures. Among the many ways of modify slightly the LJ potential, we have used the following form:

$$V(r)=V_0+V_1+V_2+V_3 \quad (1)$$

where:

$V_0=A_0[1/r^{12} - 2/r^6+1]$      if   $r<1$
$V_1=[(r-1)^2(r+1-2c)^2 / (c-1)^4] - 1$      if   $r<c$
$V_2= - A_2[(r-d_2-s_2)^2 (r-d_2+s_2)^2] / s_2^4$      if   $d_2-s_2<r<d_2+s_2$
$V_3= A_3[(r-d_3-s_3)^2 (r-d_3+s_3)^2] / s_3^4$      if   $d_3-s_3<r<d_3+s_3$      (0 otherwise in all cases)

$V_0$ is the repulsive part of a LJ potential and its weight in the total potential is measured by the parameter $A_0$. The quartic term $V_1$ contributes with an attractive well to the total potential. The last two terms are fine tuning terms that provide a small minimum of amplitude $A_2$ centered at $d_2$, and a small maximum of amplitude $A_3$ centered at $d_3$. They were adjusted to penalize appropriately the triangular lattice, and/or favoring the martensitically related structure. In consequence, the potential is fully determined by the set of parameters $P=\{A_0, A_2, A_3, c, d_2, s_2, d_3, s_3\}$. See in figure 2(a) the contribution of each term to the total potential energy for a given set of parameters, where $E_i$ is the potential energy related to the term $V_i$.

The finding of this family of potentials having different possible ground states and a controlled way to switch their relative stability is the main point of our paper. We claim that such potentials are useful benchmarks to study the properties of martensitic transformations in detail. In the rest of the paper we start such a study.

We present two parameter sets $P_1$ and $P_2$ which drive a triangular-rombohedral transition (T-R) and a triangular-square transition (T-S) respectively:

$P_1$ = {$A_0$, 0.003, 0.01, 1.722, 0.98, 0.04, 1.74, 0.2} with variable $A_0$, for the T-R transformation. The transition value is $A_0^c$ = 0.067.

$P_2$ = {0.024, $A_2$, 0.01, 1.730, 0.98, 0.1, 1.74, 0.2} with variable $A_2$, for the T-S transformation. The transition value is $A_2^c$= 0.022.

For each set of parameters, we calculate the energy of a structure characterized by the two distances a and b defined in figure 1(a). A triangular structure is obtained when a=b, whereas a square structure is obtained for a=sqrt(3)b. Values of the ratio a/b in between correspond to rombohedral structures.

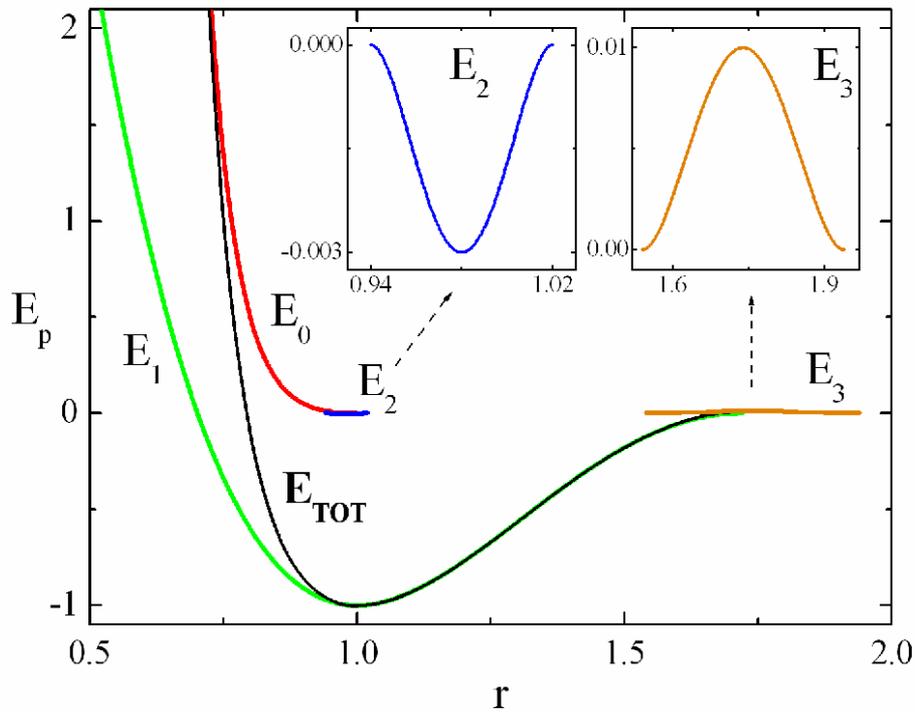

Figure 2: Potential energy as a function of the interparticle distance for the set $P_1$ and $A_0$=0.063. Red ($E_0$), green ($E_1$), blue ($E_2$) and orange ($E_3$) curves are the potential energy contributions to the total energy $E_{TOT}$, which is plotted in black.

For the set $P_1$ with a parameter $A_0$=$A_0^c$=0.067 the contour energy plot presented in figure 3(a) shows the equal stability of the triangular and a rombohedral phase. When $A_0$>$A_0^c$ ($A_0$<$A_0^c$) the energy minimum corresponding to the triangular (rombohedral) structure becomes more stable.

For the potential set $P_2$, the transition is driven by the parameter $A_2$ and the critical value is $A_2^c$= 0.022. Energy level curves (figure 3(b)) show the two minima at the triangular and the square configuration. Increasing (decreasing) $A_2$ stabilizes the triangular (square) structure.

Note that although we will speak of the configuration found as the ground state (or lower energy configurations) in our system, we have performed our search among a set of structures that of course do not exhaust the whole possibilities. Anyway, the results of the numerical implementation presented in the following sections allows to claim that if the structures presented are not the true ground state of the system, we have a least a metastability range in which they are local minima.

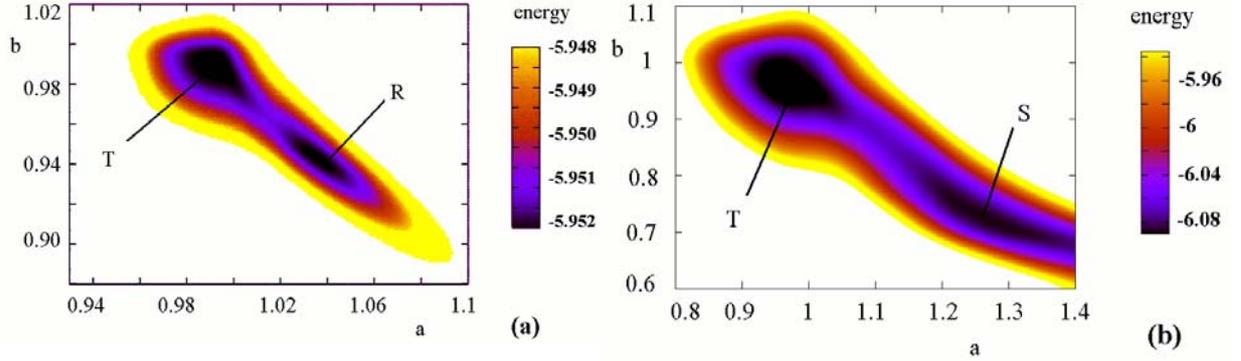

Figure 3: Energy levels as a function of the lattice parameters. (a) T-R transition. The two minima are in a=1.045, b=0.932 and a=b=0.994, and have an energy E=-5.953. (b) T-S transition. The two minima are in a=1.258, b=0.727 and a=b=0.965, with E=-6.087.

By mapping the position and depth of the energy minima, we obtain the plots of figure 4. We see the linear crossing of the energy curves of both phases, indicating a first order transition, and the corresponding values of the lattice parameters.

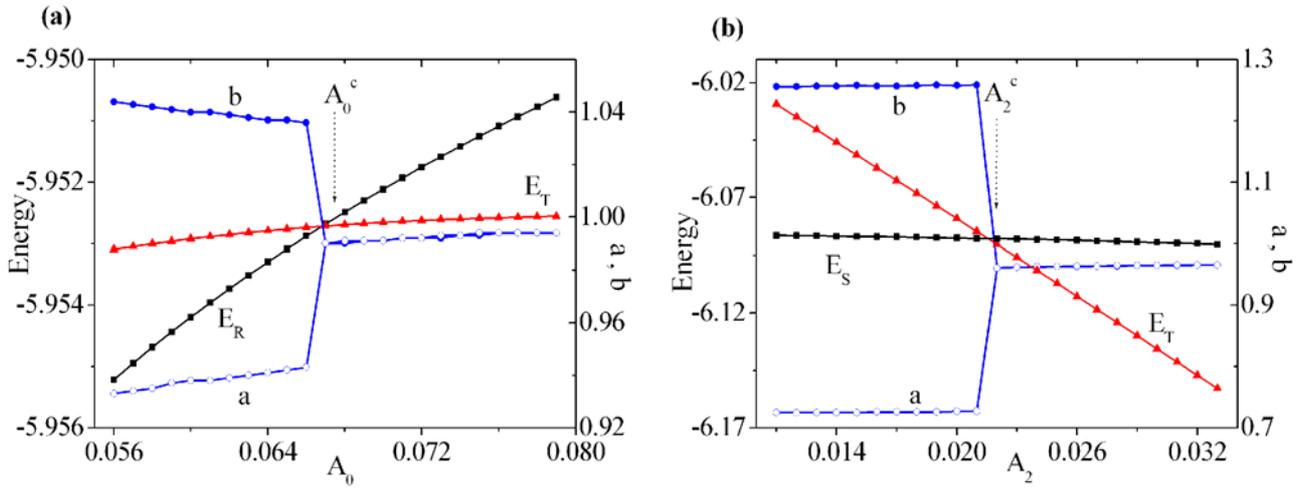

Figure 4: Lattice parameters a and b as defined in figure 1(a), and energy of the triangular ($E_T$), rombohedral ($E_R$) and square ($E_S$) structures as a function of the parameters driving the transition. (a) T-R transition. (b) T-S transition.

These two transitions involve a very small change of volume. As can be easily calculated from the lattice parameters of figure 4, the change in volume is less that one per cent for both T-R and T-S transitions. This is an important fact since actual martensitic transformations have specific volume differences between austenite and martensite that are typically in this range, too. In order to numerically study the T-R and the T-S transformation in 2D, we solve for the time dependence of the particle coordinates according to the Verlet scheme:

$$\mathbf{r}(t) = 2\mathbf{r}(t-dt) - \mathbf{r}(t-2dt) + dt^2 \, [\, \mathbf{f}(t-dt) - \mu \, \mathbf{v}(t-dt) \,] \qquad (2)$$

$$\mathbf{v}(t) = [\, \mathbf{r}(t) - \mathbf{r}(t-dt) \,] / dt$$

with $\mathbf{r}(t)$ the position of each particle, $\mathbf{v}(t)$ its velocity and $\mathbf{f}(t)$ the total force acting over the particle at time t. In all this paper we use $\mu=6$ and $dt=0.01$.

Note that we introduce a friction term proportional to the velocities, however, we do not incorporate any stochastic forcing term. This approach corresponds then to a zero temperature Langevin simulation [13]. The convenience of this approach becomes clear when it is realized that the conversion between martensite and austenite occurs in general at values of the control parameter that are not precisely the equilibrium ones (see below). This means that there is some thermal energy that is generated during the transition, and this has to be taken out of the sample to avoid generating very large particle velocities. The local friction term efficiently accomplishes this goal, whereas protocols depending on the global energy (the Nose-Hoover thermostat, for instance) would not be effective [14].

We typically prepare our starting sample by placing the particles in a monocrystalline initial configuration of a given shape, and choose parameters in Eq. (1) such that the T structure is the stable one. We relax this configuration through the dynamical algorithm of Eq. (2). At the end of this stage we obtain the relaxed starting configuration. Then the appropriate parameter is changed in the interaction potentials (either smoothly or abruptly) towards a parameter region where the competing phase is the stable one, and monitor the dynamical evolution of the system.

In order to study the memory effect, we also need to apply a mechanical stretching on two opposite ends of the sample. This is done in a very direct way, just controlling the longitudinal positions of a group of particles in some superficial region of the sample at the two sample ends, and moving these coordinates according to the external forcing conditions. Note that the transversal coordinates of these particles are left to relax according to their normal dynamical evolution, to avoid spurious strain accumulations in the surface sample region.

For quantitative evaluation and visualization purposes, we will need to determine whether some particle in the sample corresponds to the triangular (austenite) phase, or to any of the martensite variants (rombohedral, or square depending on the case[1]). We do this by calculating the distance d between neighbor particles, and drawing this segment in color when its value lies in the interval 1<d<1.3 for the T-R transition and d<0.91 for the T-S one. This will identify martensitic variants (where some segment will be displayed) against the austenite structure (where no interparticle segment will be displayed). Also, to discriminate between different variants the color of the interparticle segment will be different according to which side of the original triangular structure it corresponds. Then, we use three different colours for the three possible different orientations, called $v_1$, $v_2$ and $v_3$. Undistorted regions of the sample have no segments added. An example of such construction can be observed in the next section (see, for example, figure 5).

All the results showed in this paper correspond to a system of N=40000 particles with open boundary conditions.

## 3. The T-R transformation of a rectangular single crystal sample

To reproduce the basics of martensitic transformations using our potential, we prepared a starting single crystal sample of rectangular shape in a triangular structure and for the set of parameters $P_1$ with $A_0=0.08$.

After relaxing the original sample at the starting parameters, we change abruptly the parameter $A_0$ to the value $A_0=0.057$, where the R structure is the stable one. A time sequence of the transformation is plotted in figure 5. As can be observed in the first image of the sequence (figure 5(a)), the sample has for construction two different kind of surfaces: the (10) and (12) ones, in the basis defined by the vectors a and b of figure 1(a). The horizontal (10) surfaces are formed by a unique row of particles, whereas the vertical (12) ones consist of particles in a zig-zag configuration.

The energy of the (12) surface is larger than the (10), and this produces the nucleation of the martensitic transformation preferently at these surfaces, as was also observed in Ref. [5]. Transformation starts by the nucleation of martensite variants that invade the interior of the sample during the transformation process. In this progression to the interior, it is also apparent the existence of two variant wedge sectors that are also ubiquitously observed in real samples. The borders of the wedge correspond to the habit lines of the transformation. In the final configuration of figure 5 (d) the existence of the wedges that were important during the transformation has practically disappeared. At this final stage the existence of twinned martensite is also appreciable.

---

[1] For the T-R transformation, the fact that the symmetry group of T contains that of R allows us to define T as austenite, and R as the martensite phase. For the T-S transformation, such a univocal choice is not possible, since none of the two symmetry groups contains the other one. Nevertheless, for clarity, we continue to identify the T structure with the austenite (and use it as the starting configuration) and the S structure as martensite.

The time scale of the transformation process, as well as the typical grain size obtained, are very dependent on the depth of the quench, namely, how far the driving parameter is from the point of equilibrium between austenite and martensite. Deep quenches produce a more rapid transformation, and at the same time a much finer martensitic texture.

In figure 6 we can see a time sequence corresponding to an $A_0$ value closer to the transition: Starting from the same initial condition as in figure 5, but setting $A_0=0.063$ (note that the equilibrium for the set $P_1$ is reached when $A_0^c=0.067$), the transformation proceeds at a lower rate by forming a single wedge that progressively invades most of the sample. Note that the final state has just a few big grains of martensitic variants. As a consequence, the borders are more distorted than in the previous case.

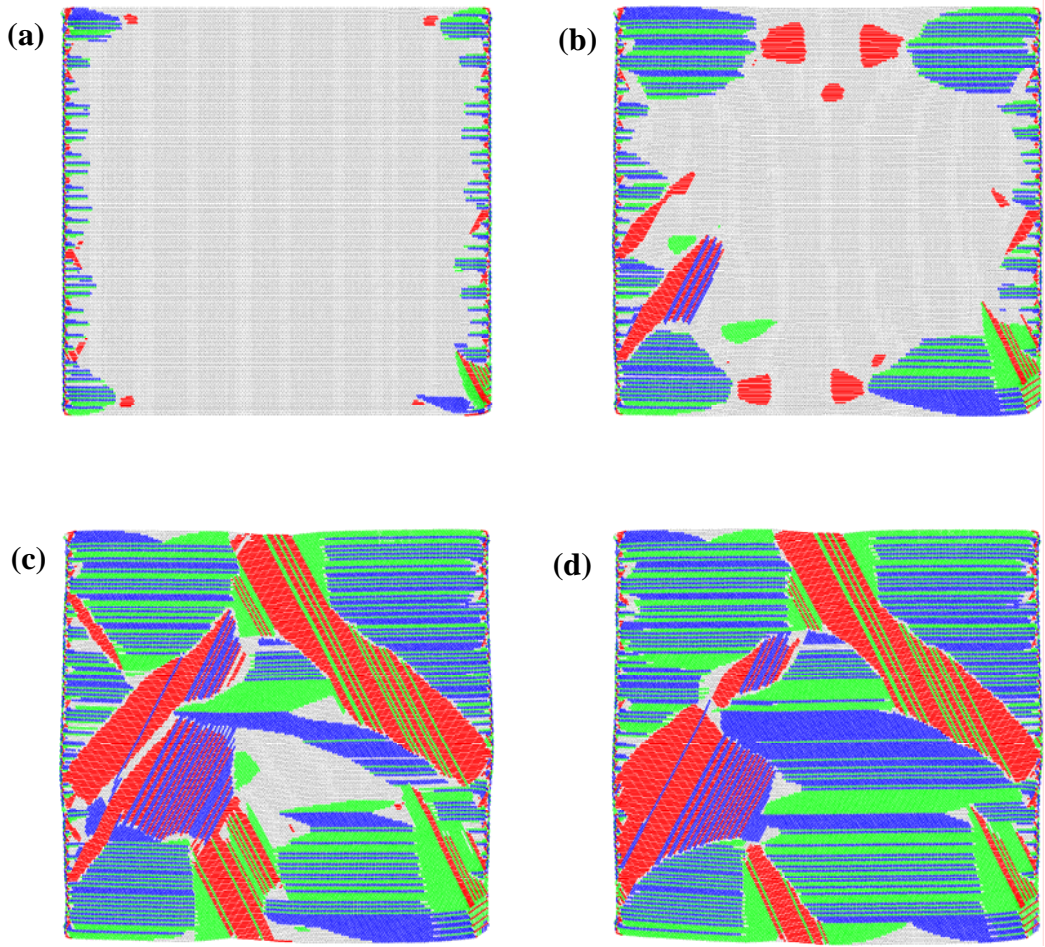

Figure 5: Time sequence for the transition T-R with $A_0=0.057$. Snapshots correspond to times (a) $t=1\times10^4$, (b) $t=5\times10^5$, (c) $t=1.5\times10^6$ and (d) $t=1\times10^7$.

The final state of figure 6 is obtained after five times the number of steps needed to obtain the final structure of figure 5. This result is quantified in figure 7(a), where we plot the number of grains in the final state and the number of steps needed to reach the final state as a function of $A_0$. We also observe a systematic increase in grain size as the final value of $A_0$ is placed closer to the transition value, reflecting the fact that the transition occurs in conditions progressively closer to thermodynamic equilibrium. This is also consistent with the fact that the transition times increase correspondingly. In figure 7(b) we plot the difference between the bulk energy (i.e., the energy of the system without the contribution of the particles located near the boundaries) of the final configuration, and the ideal energy of a single variant martensite at the same value of $A_0$. This difference is a measurement of the excess energy associated to grain boundaries between different variants and elastic energy accumulated in the sample. The increasing of this value when $A_0$ is reduced is compatible with the corresponding increase in number of grains and with the amount of grain boundaries present.

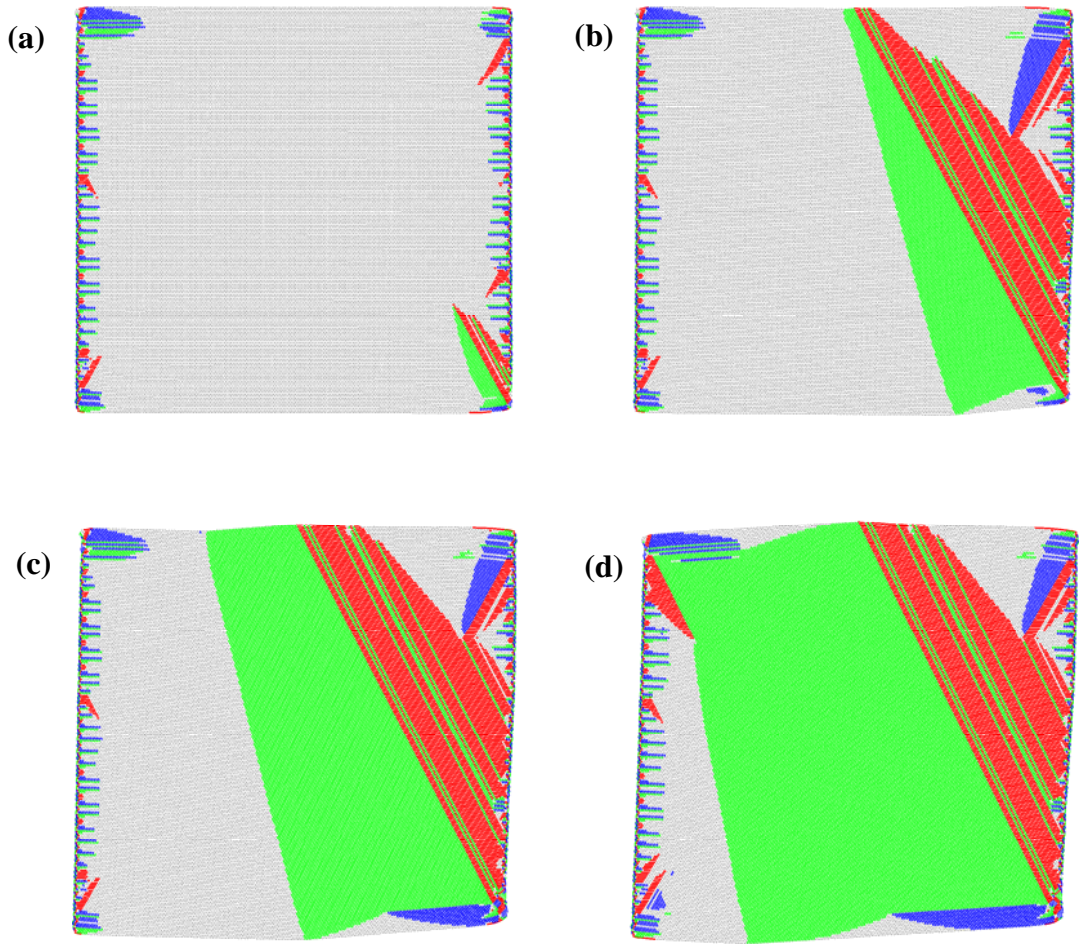

Figure 6: Time sequence for the transition T-R with $A_0$=0.063. Snapshots correspond to times (a) $t=5 \times 10^5$, (b) $t=1 \times 10^7$, (c) $t=2 \times 10^7$ and (d) $t=5 \times 10^7$.

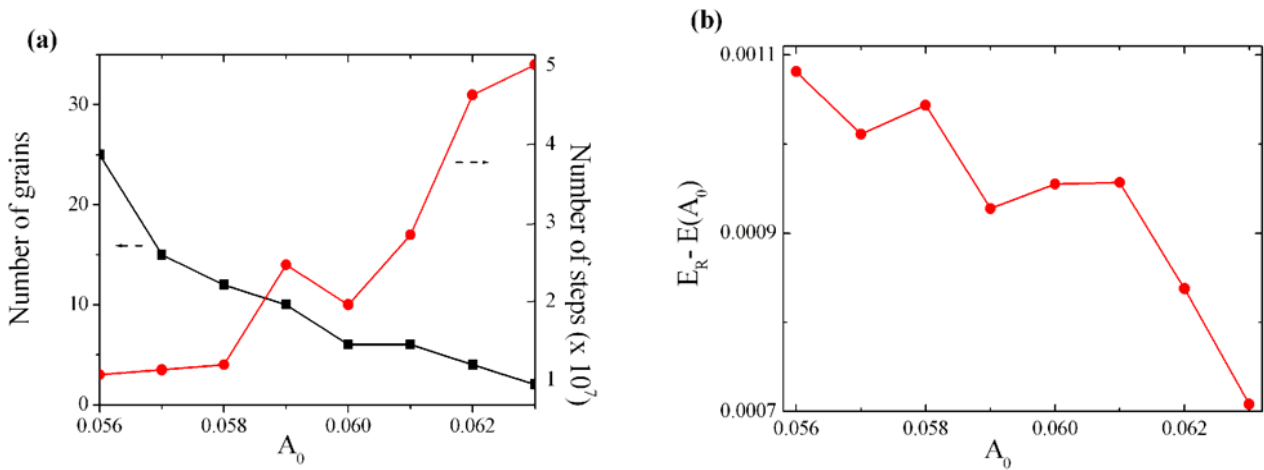

Figure 7: (a) Number of grains in the final state (black square symbols) and the number of steps (x$10^7$) needed to reach the final state (red circles) as a function of $A_0$. (b) Energy difference between the calculated rombohedral energy $E_R$ (plotted in figure 2) and the final energy obtained in runs with different values of $A_0$.

As the next step, we inverted the conditions of the transformation, placing back the value of $A_0$ in a region of stability of the austenite, and observed how the system retransforms back. In figure 8 we show two stages of such process for the run showed in figure 5. We found that the initial crystalline arrangement is recovered. This result is a consequence of the way in which the T-R transformation proceeds: the displacement of the particles is much lower than the interparticle distance itself during the process, and they have a unique path to come back to the initial position. As will be seen in the next section, this behavior is the main cause of the shape memory effect.

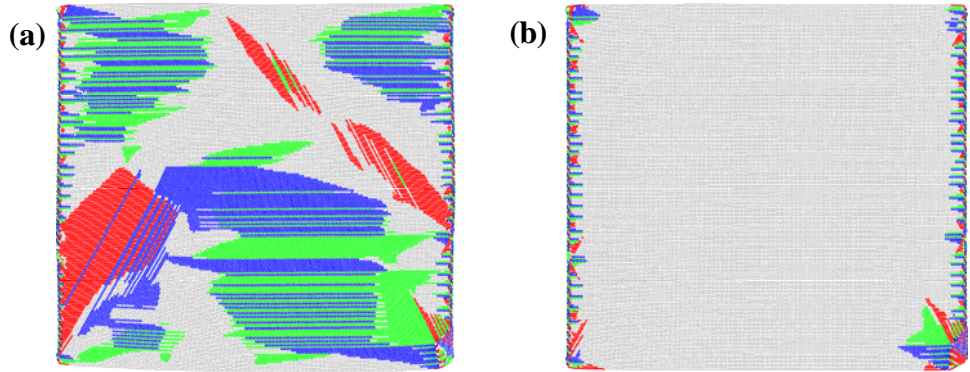

Figure 8: Retransformation of the R structure by reversing the value of $A_0$ after the T-R transition showed in figure 5. (a) Image of the structure after $5 \times 10^4$ time steps since the begining of retransformation process. (b) Final state after $1 \times 10^6$ time steps, to be compared with the initial condition (figure 5 (a)).

## 4. Shape memory effect in the T-R transformation

Shape memory effect is one of the most characteristic effects observed in martensitic transformations. Although the effect for macroscopic samples applies to rather general shape changes, the essential features of the effect are discernible by describing the uniform shape change of a small piece of material, and we concentrate in this case.

Consider the T-R transformation as described in the previous section. We saw that the original rectangular sample transforms to a martensite phase, formed by a collection of crystal of different variants. The shape of the sample remains macroscopically the same (except for some rippling of the surface) in this transformation. As we saw in the previous section, upon a reversion of the parameters the sample transforms back to the original single crystalline configuration. More than that, it can be seen that most particles return to the same initial position, having in particular the same set of neighbors. This occurs because, starting from the R structure, each particle has a unique way to return to the parent T structure. The same description applies if, with the sample in the (polycrystalline) martensite phase, we apply a mechanical distortion that produces a macroscopic shape change in the sample. Microscopically, this shape change is typically accommodated by a reconversion among different variants, but still, each particle in the system has a unique way to return to its original position in the T structure. This means that when parameters change and the sample transforms back to the T structure, the original shape is recovered. This is the essence of shape memory effect.

We performed the protocol of transformation-deformation-relaxation-retransformation in our system to see to what extent the ideal description of the shape memory effect applies to our model. Starting from the final state of figure 5(d), we deformed the sample in the x-direction, as shown in figure 9. A reconversion of variants is clearly observed. In figure 9(d) almost all the sample is in the martensitic variant that most easily accommodates the deformation. Note the macroscopic change of shape that occurred in the process. Most importantly, if now the external force is withdrawn, the system maintains the deformed state (we call *relaxation* to this stage of the process). If then the parameter $A_0$ is changed to a value in which the austenite phase is stable ($A_0=0.08$) we observe that the retrasformation proceeds until the system recovers the monocrystaline triangular structure of the initial condition and its original shape (see figure 10).

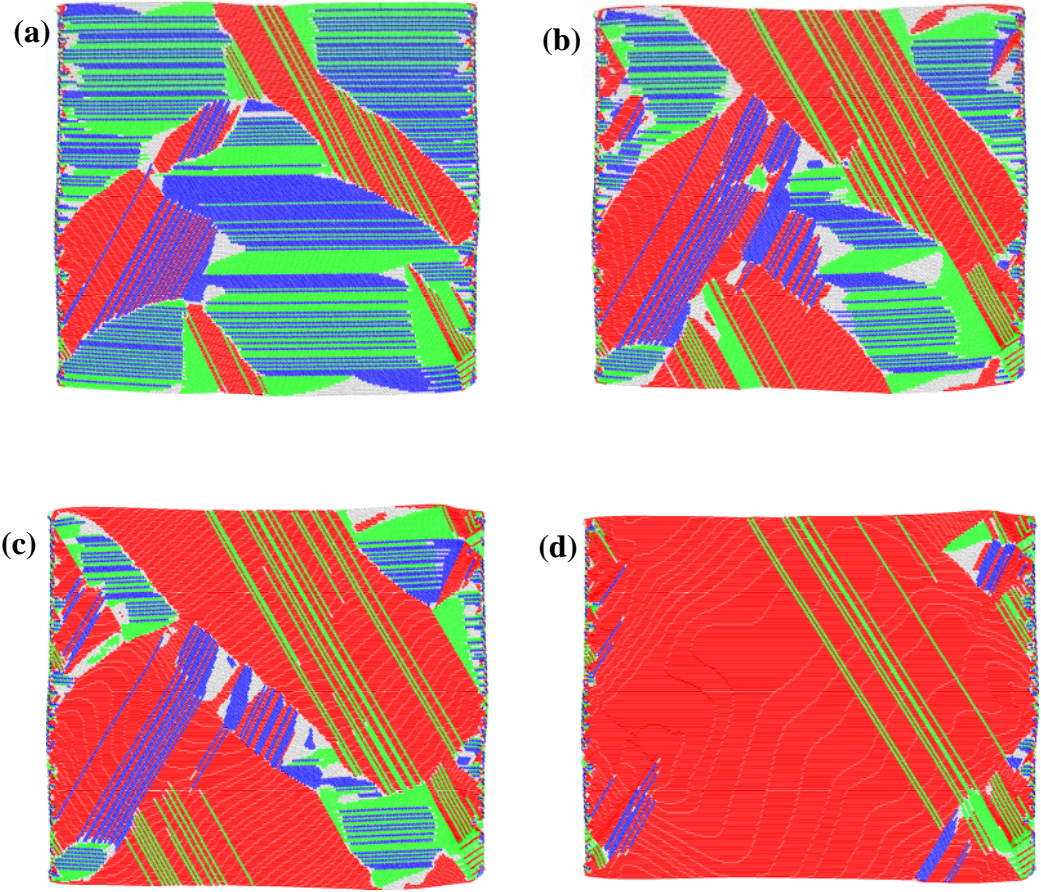

Figure 9: Stretching along the x direction, starting from the final state of figure 5. (a) Image of the structure after $5 \times 10^4$ time steps since the begining of streching process. (b) Imagen after $3 \times 10^6$ time steps. (c) Imagen after $3.5 \times 10^6$ time steps. (d) Imagen after $5.5 \times 10^6$ time steps.

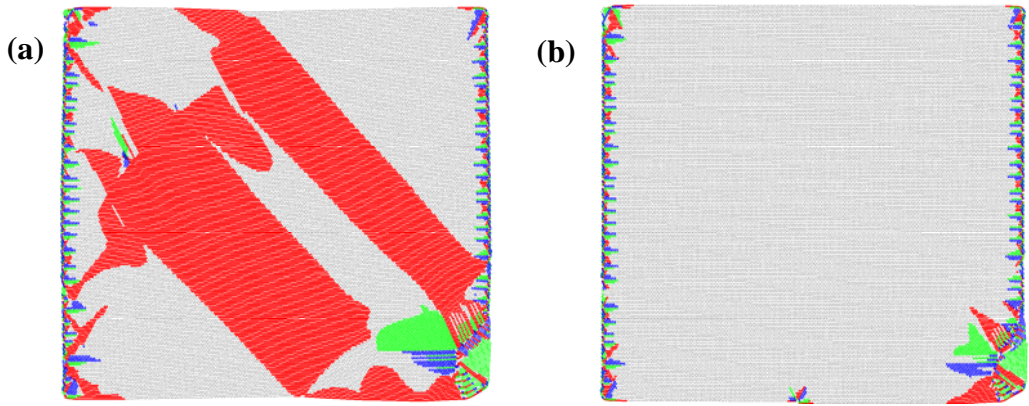

Figure 10 : Retransformation process, starting from the final state of figure 9 and increasing $A_0$ with no applied force (a) Intermediate stage after $3 \times 10^6$ steps since the begining of the retransformation process. (b) Final monocrystaline state after $4 \times 10^6$ steps.

The full cycling protocol can be conveniently described by the evolution in time of two quantities giving complementary information. One is the fraction of particles of the system in each martensitic variant, $\phi_i = Nv_i/N$, with $Nv_i$ the number of particles belonging to the martensitic variant $v_i$ and $N$ the total number of particles. The other quantity measures the overall aspect ratio of the sample. This parameter is defined as the ratio $I_x/I_y$, with $I_x = \langle (x(i) - \langle x \rangle)^2 \rangle$ and $I_y = \langle (y(i) - \langle y \rangle)^2 \rangle$, and where brackets note averages over particles in the system, labeled by i. The evolution of these quantities is displayed in figure 11. The first stage corresponds to the T-R transformation of figure 5. The fraction of variants of figure 11(a) have a quick growth and then a slower approach to the final value, whereas the aspect ratio showed in figure 11(b) reflects a slight and realization dependent shape change suffered by the system in the transformation. Stage II is the deformation process. In panel (a) the reconversion of variants is clearly observed: the energetically favored variant $v_1$ grows whereas the other two decrease. This reconversion is also accompanied by a notable change of shape measured by the aspect ratio in panel (b). During the third stage the system relaxes with no forces applied. The fraction of variants remains almost constant and the aspect ratio converges to a value lower than the maximum but still much higher than the initial one. In the last stage of retransformation the system returns to the austenite phase, and this is reflected in the vanishing of martensitic phase. The shape recovery in turn, manifests in the return of $I_x/I_y$ to its starting value.

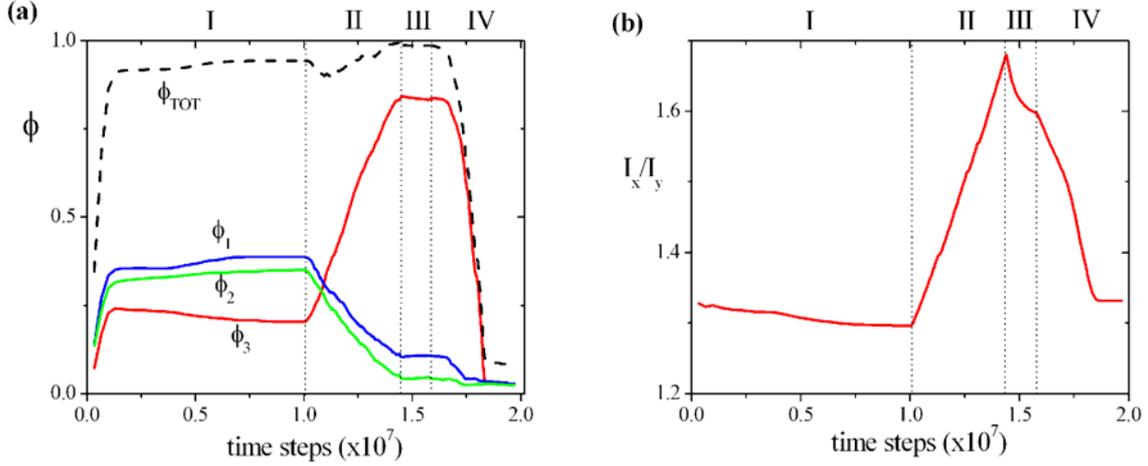

Figure 11: (a) Evolution of the fractions of each martensite variant $\phi_i$ and its sum $\phi_{TOT}=\phi_1+\phi_2+\phi_3$. (b) Evolution of the ratio of moments of inertia $I_x/I_y$. The stages of the process are: T-R transformation (I), deformation (II), relaxation (III) and retransformation (IV).

## 5. Absence of shape memory effect in the T-S transformation

Contrary to the case of the T-R transformation, in the case of the T-S transition, particles in the S phase do not have a unique transformation back to the austenite phase. In a formal context, this fact has been formulated as the statement that in this case, the symmetry group of the martensite is not a sub-group of the symmetry group of the austenite [9].

In any of the two forms, this fact is enough to explain the absence of shape memory effect in the T-S transformation, and we now show results in our model that are consistent with this prediction.

We start with a set of parameters in which the triangular phase is stable ($A_2$ initial = 0.030) and change $A_2$ to the value $A_2=0.013$, for which the square phase has lower energy. After the T-S transformation is completed, the state of figure 12(a) is obtained. We observe a sample in a square martensitic state of policrystaline structure formed by grains of square phase in the three possible directions of deformation (see figure 1(a)). Reversing the parameter $A_2$ to the initial value and letting the system evolve, the configuration of figure 12(b) is obtained. The final state is a polycrystalline triangular structure retaining some small grains of square phase. This defective structure has to be compared with the almost perfect recovery of the original sample observe for the case of the T-R transition (figure 8).

If from the configuration of figure 12(a) we apply a mechanical stretching, and then allow the sample to relax under no external force, we obtain the result showed in figure 13(a). It is seen that, as in the T-R case, there has been a reconvertion between variants compatible with the applied stretching, and the sample has acquired an appreciable global deformation. Reverting now the parameter $A_2$ to a value in which the T phase is the stable one, we obtain the configuration in figure 13(b). A polycrystalline T phase is obtained. The evolution of the fraction of particles in each variant and the ratio of moments of inertia during this transformation-deformation-retransformation protocol can be observed in figure 14.

In addition to the fact that now two variants are favored during the stretching process, the crucial point is that the sample does not recover its original macroscopic shape, but retains the deformation that was applied in the martensitic state (compare figure 14(b) with figure 11(b) obtained for the T-R transformation). This result clearly manifests the absence of the shape memory effect in the T-S transformation.

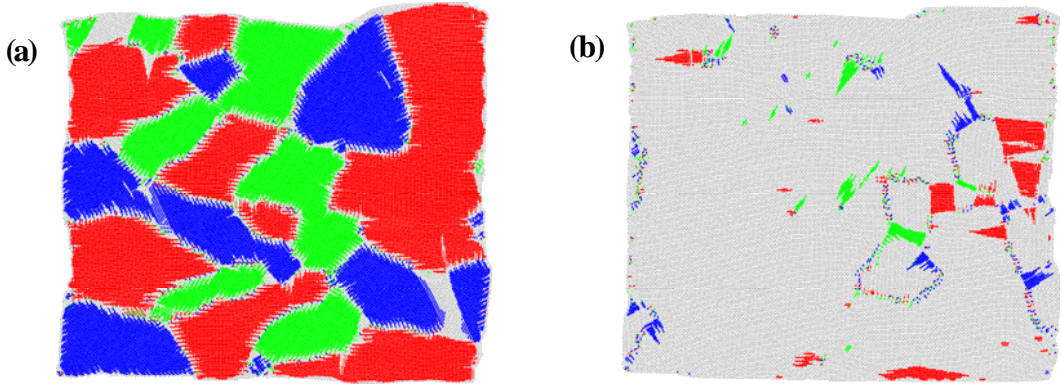

Figure 12: Final states. (a) T-S transformation (after t=4x10⁶ time steps). (b) Image of the system after 2x10⁶ steps since the begining of the retransformation process.

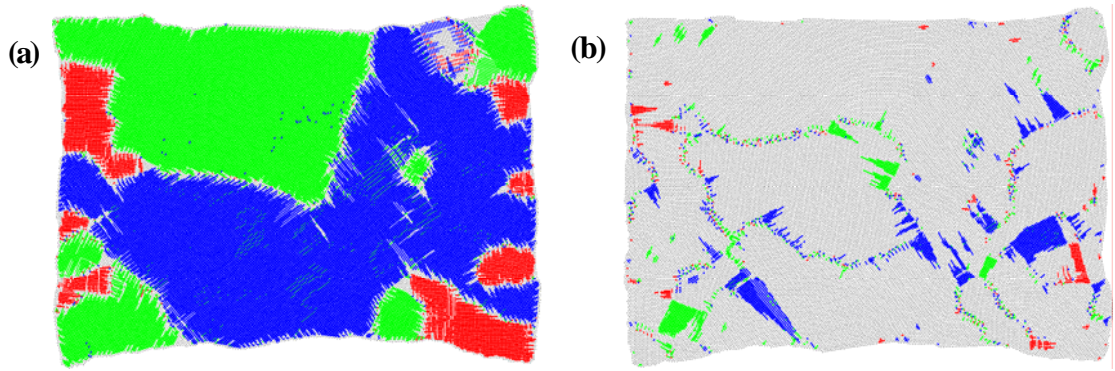

Figure 13: (a) State after the mechanical stretching and relaxation (with no applied forces) of the sample of Fig. 12(a) For this picture t=3x10⁷ time steps. (b) State after retransformation, at t=4x10⁷ time steps. Grains of differently oriented triangular phases are observed.

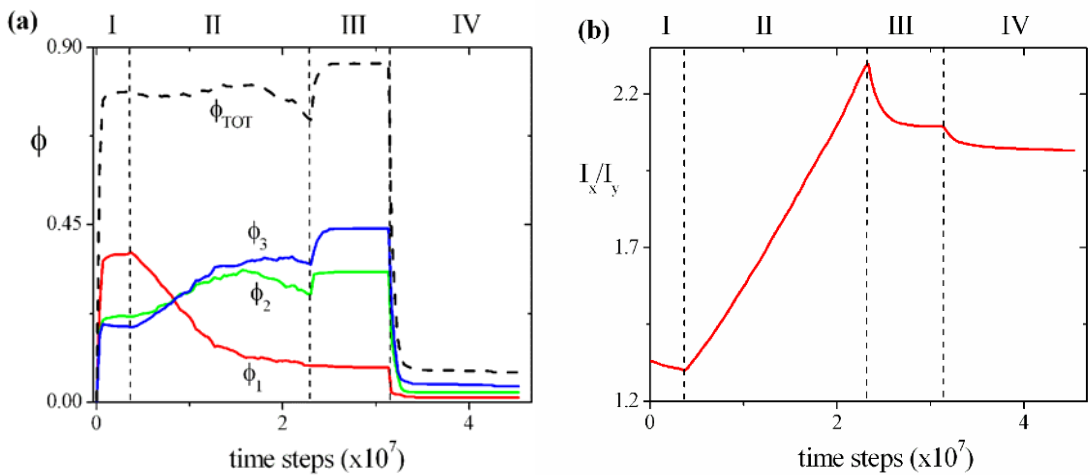

Figure 14: (a) Evolution of the fractions of each martensite variant $\phi_i$ and its sum $\phi_{TOT}=\phi_1+\phi_2+\phi_3$. (b) Evolution of the ratio of moments of inertia $I_x/I_y$. The stages of the process are: T-S transformation (I), deformation (II), relaxation (III) and retransformation (IV).

## 6. Conclusions

We have presented a two body isotropic potential that produces different crystalline configurations when some parameter is varied. Upon such a variation of parameters the stability of different crystalline configuration can change relatively to each other, and this can produce a martensitic transformation in the system. We have studied in two dimensions the cases of a triangular-rombohedral transformation and a triangular-square transformation. For the T-R case, we have observed in detail how the transformation and retransformation proceed in time, and also verified that the system displays the shape memory effect associated to some martensitic transformations. For the case of the T-S transformation, the same protocols show again the existence of a martensitic transformation, but the shape memory effect is absent, in accordance with theoretical expectations.

The present and related work [8] demonstrate that complex interatomic potentials in multi-species systems are not necessary to study the basics of the phenomena associated with martensitic transformations. We have also corroborated in a direct way that thermal fluctuations are not crucial for the transformation, at least as far as it can be described by our model.

We plan to study in the near future more realistic cases as for instance some kind of three dimensional transformation for which we already have appropriate interaction potentials, as well as other interesting effects in this kind of systems, in particular, the superelasticity effect and the possibility of the two-way shape memory effect.